\documentclass{iopjournal_mod}
\usepackage{amsfonts}
\usepackage{amsmath}
\usepackage{subcaption}
\usepackage{longtable}
\usepackage{booktabs}
\usepackage{array}
\usepackage{makecell}
\usepackage{etoolbox}
\makeatletter
\patchcmd{\@maketitle}{\@top\@journalinfo}{}{}{}
\makeatother

\usepackage{ragged2e}

\begin{document}
\justifying
\articletype{Preprint}
%%%%%%%%%%%%%%%%%%%%%%%%%%%%%%
\title{Three-dimensional Fundamental Diagrams of Five-neighbor Particle Cellular Automata}

\author{Kazuya Okamoto$^1$\orcid{0009-0008-2638-7780} and Daisuke Takahashi$^2$\orcid{0009-0005-8998-1152}}

\affil{$^1$Faculty of Information Design, Heisei International University, Saitama, Japan}

\affil{$^2$Department of Pure and Applied Mathematics, Waseda University, Tokyo, Japan}

\email{okamoto@hiu.ac.jp}

%%%%%%%%%%%%%%%%%%%%%%%%%%%%%%
\begin{abstract}
%%%%%%%%%%%%%%%%%%%%%%%%%%%%%%
\justifying
We analyze five-neighbor particle cellular automata whose conventional two-dimensional fundamental diagrams are multivalued, but whose mean flow is uniquely determined by introducing a second density. 
We first consider binary rules for which the second density is conserved, and then examine rules for which the second density is not conserved but converges asymptotically. 
These examples give three-dimensional fundamental diagrams in which the mean flow is determined by the particle density and the second density.

We then investigate whether this single-valued structure is preserved under real-valued max-plus extensions. 
There are some rules where two different max-plus extensions are introduced, and numerical simulations show that both extensions preserve the same single-valued three-dimensional fundamental diagram. 
These observations imply that, in constructing real-valued max-plus extensions, it is important to choose the flux function and the second density consistently.
\end{abstract}
%%%%%%%%%%%%%%%%%%%%%%%%%%%%%%
\section{Introduction}
%%%%%%%%%%%%%%%%%%%%%%%%%%%%%%
Cellular automata (CA) are dynamical systems in which time, space, and the state variable are fully discretized. Since the introduction of Conway's Game of Life in the 1970s \cite{gardner1970} and Wolfram's classification of one-dimensional elementary cellular automata (ECA) in the 1980s \cite{wolfram1983}, CA have been widely used as simple and powerful models for describing complex systems, self-organization, and nonlinear phenomena \cite{wolfram2018}.

Although CA and differential equations have traditionally been regarded as distinct mathematical frameworks, the introduction of ultradiscretization \cite{tokihiro1996} established a rigorous connection between them through a limiting procedure. A notable example is an elementary cellular automaton of rule number 184 (ECA184), which is a fundamental CA model for traffic flow. Nishinari and Takahashi showed that an ultradiscrete Burgers equation can be derived from a discrete Burgers equation through ultradiscretization and that, under a suitable parameter choice, it reduces to ECA184 \cite{nishinari1998}. This result provides a direct mathematical link between a nonlinear equation related to fluid dynamics and a cellular automaton used in traffic flow modeling.

In the traffic-flow interpretation of ECA184, the state value 1 represents a particle and the state value 0 represents an empty site. A particle moves only when the site in front of it is empty, and the total number of particles is conserved during time evolution. Cellular automata with such a conservation property are called particle cellular automata (PCA). Since the time evolution rule of a PCA can be written in the conservation form using a local flux function \cite{hattori1991}, the particle density, defined as the spatial average of the state variable, is conserved. The relation between this density and the average flow is called the fundamental diagram, and it plays a central role in characterizing the macroscopic behavior of particle systems.

The time evolution equations obtained through ultradiscretization are expressed in terms of max or min operations and ordinary addition. Such equations can be represented by max-plus or min-plus algebraic systems, where ordinary addition and multiplication are replaced by the maximum (or minimum) operation and addition, respectively. Nishinari and Takahashi introduced an ultradiscrete Cole--Hopf transformation for the ultradiscrete Burgers equation, which reduces it to an ultradiscrete diffusion equation. The time evolution of the ultradiscrete diffusion equation enables the analysis of the asymptotic behavior of the Burgers cellular automaton. In particular, under periodic boundary conditions, the solution eventually reaches a stable pattern, and the mean flow of particles is determined solely by the particle density \cite{nishinari1998}. Since the state variable can be real values, max-plus and min-plus formulations are useful not only for describing binary CA rules but also for analyzing the behavior of multivalued particle systems. In traffic flow modeling, min-plus formulation has also been used to derive density-flow relationships analytically by solving the corresponding eigenvalue problems \cite{lotito2005,farhi2011}.

For three-neighbor PCA, after excluding rules reducible to fewer-neighbor systems and grouping rules related by space reflection and Boolean conjugation symmetries, there remains only one nontrivial equivalence group of rules including ECA184. We denote ECA184 by PCA3 and use it as the canonical representative of this equivalence group of three-neighbor PCA rules. For this system, the mean flow is uniquely determined by the particle density. Similar single-valued fundamental diagrams are known for several four-neighbor PCAs \cite{takahashi2011}. However, there exist PCA rules for which the relation between density and mean flow is not single-valued. The multivaluedness becomes more serious for five-neighbor PCA (PCA5). Since the neighborhood is larger, asymptotic behaviors of PCA5 exhibit a rich variety. Some PCA5 rules have single-valued fundamental diagrams and have recently been used as mathematical models for pedestrian flow \cite{okamoto2025}. On the other hand, many PCA5 rules have several possible values of the mean flow for the same density.

The conventional two-dimensional fundamental diagram is not sufficient for describing PCA5 in general from the viewpoint of exact analysis of asymptotic behavior \cite{okumura2013}. To overcome this difficulty, Endo and Takahashi proposed a three-dimensional fundamental diagram for PCA5 with two conserved densities \cite{endo2022}. In their framework, an additional conserved density, such as the density of a specific local pattern, is introduced as a second independent density. The mean flow is then uniquely determined by the pair of the particle density and the additional conserved density. This result shows that the apparent multivaluedness of the conventional density-flow relation can be resolved by introducing an appropriate density.

In this paper, we present several binary PCA5 rules where three-dimensional fundamental diagrams can be constructed by introducing an additional density. For this purpose, we use either a second conserved density or an asymptotic second density.  Moreover, we formulate the corresponding time evolution equations in max-plus algebraic form and introduce real-valued extensions of several binary PCA5 models. We then examine how the behavior of the extended systems can be characterized within the framework of three-dimensional fundamental diagrams.

It is therefore natural to ask whether the single-valued structure of the three-dimensional fundamental diagram is preserved under real-valued max-plus extensions. If this property survives beyond the binary setting, it may provide a unified viewpoint connecting binary PCA and real-valued ultradiscrete systems.
%%%%%%%%%%%%%%%%%%%%%%%%%%%%%%
\section{Particle cellular automata and fundamental diagrams}
\label{sec:pca_preliminaries}
%%%%%%%%%%%%%%%%%%%%%%%%%%%%%%
In this section, we introduce particle cellular automata and the basic quantities used in this paper. First, one-dimensional binary cellular automata and particle cellular automata are defined. Then, the naming system for PCA5 rules is explained \cite{okumura2013}. Finally, we define density, flux, mean flow, and fundamental diagrams.
%%%%%%%%%%%%%%%%%%%%%%%%%%%%%%
\subsection{Binary cellular automata}
%%%%%%%%%%%%%%%%%%%%%%%%%%%%%%
A cellular automaton is a discrete dynamical system in which space, time, and state variable are all discrete. Let us consider one-dimensional binary cellular automata. The spatial index is denoted by $j\in\mathbb{Z}$, and the time step is denoted by $n\in\mathbb{Z}_{\geq 0}$. The state variable at site $j$ and time $n$ is denoted by $u_j^n$, and it takes either 0 or 1.

A five-neighbor binary CA is given by
\begin{equation}
  u_j^{n+1}
  =
  f(u_{j-2}^n,u_{j-1}^n,u_j^n,u_{j+1}^n,u_{j+2}^n),
  \label{eq:ca5_rule}
\end{equation}
where
  $f:\{0,1\}^5\to\{0,1\}$
is a local rule. Since there are $2^5=32$ possible configurations, the local rule depends on the values for these configurations. Hence, the number of all five-neighbor binary CA rules is
$2^{(2^5)}=2^{32}$. The rule number of a binary CA is determined by arranging the values of $f$ for configurations in the lexicographic order 11111, 11110, 11101, 11100, $\ldots$, 00001, 00000 and interpreting the binary sequence of values as a binary integer. For example, if the values of $f$ are given by 10111000111110001000100011111000 ($f(11111)=1$, $f(11110)=0$, $f(11101)=1$, $\ldots$, $f(00000)=0$), then the corresponding rule number is $(10111000111110001000100011111000)_2$ $=3103295736$.
Thus, this rule is denoted by CA5-3103295736.

Throughout this paper, we impose a periodic boundary condition. If the system size is $K$, then
\begin{equation}
  u_{j+K}^n=u_j^n
  \label{eq:periodic_boundary}
\end{equation}
holds for any $j$ and $n$.
%%%%%%%%%%%%%%%%%%%%%%%%%%%%%%
\subsection{Particle cellular automata}
%%%%%%%%%%%%%%%%%%%%%%%%%%%%%%
A binary CA is called a particle cellular automaton, or PCA, if the number of 1's is conserved under time evolution \cite{takahashi2011}. A site with value 1 is regarded as being occupied by a particle, while a site with value 0 is regarded as empty.

For a periodic system of size $K$, the number of particles at time $n$ is $\sum_{j=1}^{K} u_j^n$. Then, a CA is a PCA if
\begin{equation}
  \sum_{j=1}^{K} u_j^{n+1}
  =
  \sum_{j=1}^{K} u_j^n
  \label{eq:particle_conservation}
\end{equation}
holds for any $n$ and any initial configuration.

For such a conservative CA, the time evolution can be written in a conservation form \cite{hattori1991}. In the five-neighbor case, we use the form
\begin{equation}
  u_j^{n+1}
  =
  u_j^n
  +
  q(u_{j-2}^n,u_{j-1}^n,u_j^n,u_{j+1}^n)
  -
  q(u_{j-1}^n,u_j^n,u_{j+1}^n,u_{j+2}^n),
  \label{eq:pca5_conservation_form}
\end{equation}
where $q(a,b,c,d)$ is called the flux. The first flux term represents an inflow from site $j-1$ to site $j$, and the second flux term represents an outflow from $j$ to $j+1$. Thus, different PCA5 rules can be specified by their different rule tables of flux $q(a,b,c,d)$. This rule table is used throughout the following sections.
%%%%%%%%%%%%%%%%%%%%%%%%%%%%%%
\subsection{Rule naming for PCA5 rules}
\label{subsec:pca5_numbering}
%%%%%%%%%%%%%%%%%%%%%%%%%%%%%%
We next recall the naming convention for PCA5 rules. As explained by Okumura et al. \cite{okumura2013}, the rules satisfying the particle conservation are extracted from all five-neighbor binary CA. Spatial reflection symmetry and Boolean conjugation symmetry are then used to identify equivalent rules. Reflection symmetry is obtained by reversing the spatial direction. If two rules are related by the transformation
\begin{equation}
  f_1(a,b,c,d,e)
  =
  f_2(e,d,c,b,a),
\end{equation}
then they are regarded as equivalent under reflection. Boolean conjugation symmetry corresponds to exchanging the roles of 0 and 1. If two rules are related by
\begin{equation}
  f_1(a,b,c,d,e)
  =
  1-f_2(1-a,1-b,1-c,1-d,1-e),
\end{equation}
then they are regarded as equivalent under Boolean conjugation.

By applying these reductions, five-neighbor PCAs are classified into equivalence groups. Okumura et al. obtained 428 five-neighbor PCA rules, reduced them to 129 groups by reflection symmetry and Boolean conjugation symmetry, and then excluded the rules reducible to PCAs of four or fewer neighborhoods. As a result, 115 independent groups remain \cite{okumura2013}.

Let us define the canonical representative of each group as the rule with the minimum rule number in the group. Then, the 115 groups can be arranged in increasing order of the canonical rule numbers and we can number the groups as 5-1, 5-2, $\ldots$, 5-115. The canonical representatives are named PCA5-1, PCA5-2, $\ldots$, PCA5-115, respectively.  Their equivalent rules in their own groups are named by adding the suffixes C, R and CR which denote conjugation, reflection and conjugate reflection, respectively. Thus, PCA5-28R means the reflected rule of PCA5-28 in the 28th group of PCA5. In this paper, we use this naming convention. The complete list of rules is shown in Table~\ref{tab:pca5_rules} of Appendix~\ref{app:pca5_rule_table}. Moreover, we can distinguish PCA rules by their fluxes. The complete list of fluxes for all PCA5 rules is shown in Table~\ref{tab:pca5_fluxes} of Appendix~\ref{app:pca5_flux_table}.
%%%%%%%%%%%%%%%%%%%%%%%%%%%%%%
\subsection{Density of particles and of local patterns}
%%%%%%%%%%%%%%%%%%%%%%%%%%%%%%
The particle density is defined by
\begin{equation}
  \rho_1
  =
  \frac{1}{K}
  \sum_{j=1}^{K} u_j^n.
  \label{eq:rho1_def}
\end{equation}
Since the system is a PCA, the number of particles is conserved, and therefore $\rho_1$ is independent of $n$. We also use densities of local patterns. For a binary word $w=x_1x_2\cdots x_m$, we define $\#w$ by the number of occurrences of local pattern $w$ in a spatial period. Its corresponding density is defined by
\begin{equation}
  \rho_w
  =
  \frac{\#w}{K}.
  \label{eq:local_pattern_density}
\end{equation}
For example,
\begin{equation}
  \rho_{011}=\frac{\#011}{K},
  \qquad
  \rho_{110}=\frac{\#110}{K}.
\end{equation}
In some cases, a local pattern density is conserved under time evolution. In the other cases, the local pattern density is not conserved but converges asymptotically. Either type of density is used in constructing three-dimensional fundamental diagrams.

We use a symbol $*$ denoting either 0 or 1. For example,
\begin{equation}
  \#1*10=\#1010+\#1110,
\end{equation}
and hence
\begin{equation}
  \rho_{1*10}
  =
  \frac{\#1010+\#1110}{K}.
\end{equation}
%%%%%%%%%%%%%%%%%%%%%%%%%%%%%%
\subsection{Flux and mean flow}
%%%%%%%%%%%%%%%%%%%%%%%%%%%%%%
For a PCA written in the conservation form \eqref{eq:pca5_conservation_form}, the flux function $q(a,b,c,d)$ represents the number of particles passing through a cell boundary per time step, depending on the local configuration $(a,b,c,d)$. At time $n$, the spatial average of the flux is defined by
\begin{equation}
  \bar{q}^{\,n}
  =
  \frac{1}{K}
  \sum_{j=1}^{K}
  q(u_j^n,u_{j+1}^n,u_{j+2}^n,u_{j+3}^n).
  \label{eq:spatial_average_flux}
\end{equation}
The system is assumed to exhibit asymptotic behavior. The mean flow $Q$ is then evaluated from the spatial average of the flux in that regime. When the asymptotic behavior is stationary, $Q$ is simply the spatial average of the flux in the stationary state. When the asymptotic behavior is periodic, $Q$ is derived by averaging the mean flux over the period.

In the plots of numerical results of this paper, we compute $Q$ after a sufficiently long transient time. For rules involving an asymptotic second density, $Q$ and the second density are evaluated in the same asymptotic regime.
%%%%%%%%%%%%%%%%%%%%%%%%%%%%%%
\subsection{Two-dimensional fundamental diagram}
%%%%%%%%%%%%%%%%%%%%%%%%%%%%%%
The conventional fundamental diagram is the relation between the density $\rho_1$ and the mean flow $Q$. In this paper, we call the plot in the plane $(\rho_1,Q)$ the two-dimensional fundamental diagram. For some PCA rules, $Q$ is uniquely determined by $\rho_1$. A typical example is the elementary traffic rule, ECA184, whose fundamental diagram is single-valued. However, for five-neighbor PCA, the same value of $\rho_1$ may lead to different values of $Q$ depending on the initial arrangement of particles. In such cases, the conventional two-dimensional fundamental diagram is multivalued and is not sufficient to characterize the macroscopic behavior.
%%%%%%%%%%%%%%%%%%%%%%%%%%%%%%
\subsection{Three-dimensional fundamental diagram}
%%%%%%%%%%%%%%%%%%%%%%%%%%%%%%
Following the framework introduced by Endo and Takahashi \cite{endo2022}, we consider a three-dimensional fundamental diagram in the space $(\rho_1,\rho_2,Q)$, where $\rho_2$ denotes an additional density. This framework provides a way to resolve the multivaluedness of the two-dimensional fundamental diagram.

The additional density $\rho_2$ may be chosen in different ways. 
In some cases, it is a conserved density, such as the density of a conserved local pattern. 
Then both $\rho_1$ and $\rho_2$ are determined by the initial configuration and remain unchanged throughout the time evolution. In the other cases, $\rho_2$ is not conserved throughout the time evolution, but is defined by its asymptotic value. 
We call such a density an asymptotic second density. 

When the points observed in the space $(\rho_1,\rho_2,Q)$ form a single-valued surface, the three-dimensional fundamental diagram provides a macroscopic description that is more informative than the conventional two-dimensional diagram in the plane $(\rho_1,Q)$. In the following section, we give some results about three-dimensional fundamental diagrams.
%%%%%%%%%%%%%%%%%%%%%%%%%%%%%%
\section{Three-dimensional fundamental diagram for binary PCA5}
\label{sec:binary_pca5_3dfd}
%%%%%%%%%%%%%%%%%%%%%%%%%%%%%%
The numerical data in this section were computed under the periodic boundary condition. For each sample, the system size $K$ was chosen uniformly at random from 5, $\ldots$, 80, and $50000$ random binary initial configurations were used. Each initial configuration was generated by choosing a Bernoulli parameter uniformly at random from $[0,1]$ and then assigning binary states independently according to that parameter. 
The system was evolved for $1000$ transient steps. 
The mean flow $Q$ was then computed from the spatial average of the flux in the asymptotic regime; when time averaging was used, the average was taken over $100$ additional steps. 
The non-conserved second densities were also evaluated in the asymptotic regime.
%%%%%%%%%%%%%%%%%%%%%%%%%%%%%%
\subsection{PCA5-17R: a case with two conserved densities}
%%%%%%%%%%%%%%%%%%%%%%%%%%%%%%
We first recall PCA5-17R, a five-neighbor particle system studied by Endo and Takahashi \cite{endo2022}. It is written in the conservation form \eqref{eq:pca5_conservation_form}, where the flux $q(a,b,c,d)$ is given in Table~\ref{tab:pca5_17_flux}.
\begin{table}[b]
\centering
\caption{Flux table of PCA5-17R.}
\label{tab:pca5_17_flux}
\begin{tabular}{c|cccccccc}
$(a,b,c,d)$ & 1111 & 1110 & 1101 & 1100 & 1011 & 1010 & 1001 & 1000 \\ \hline
$q(a,b,c,d)$ & 1 & 1 & 1 & 1 & 0 & 0 & 0 & 0
\end{tabular}

\vspace{1mm}

\begin{tabular}{c|cccccccc}
$(a,b,c,d)$ & 0111 & 0110 & 0101 & 0100 & 0011 & 0010 & 0001 & 0000 \\ \hline
$q(a,b,c,d)$ & 0 & 1 & 0 & 0 & 0 & 0 & 0 & 0
\end{tabular}
\end{table}
For this system, the particle density
\begin{equation}
  \rho_1
  =
  \frac{1}{K}\sum_{j=1}^{K} u_j^n
  =
  \frac{\#1}{K}
\end{equation}
is conserved. In addition, the number of local patterns $011$ is also conserved and its density is denoted by
\begin{equation}
  \rho_{011}
  =
  \frac{\#011}{K}.
\end{equation}
Under the periodic boundary condition, $\#011=\#110$ always holds. Thus, the density $\rho_{110}$ can be used instead of $\rho_{011}$.

Endo and Takahashi classified the asymptotic solutions into two types and derived the mean flow for each type theoretically. The mean flow is uniquely determined by the two conserved densities $\rho_1$ and $\rho_{011}$, and the three-dimensional fundamental diagram is given by
\begin{equation}
  Q
  =
  \max(2\rho_1-1,\,2\rho_{011}).
  \label{eq:pca5_17_fd}
\end{equation}
Thus, although the conventional relation between $\rho_1$ and $Q$ is not single-valued, the mean flow is uniquely determined by $(\rho_1,\rho_{011})$. 
Since \eqref{eq:pca5_17_fd} is the maximum of two linear functions, the graph is composed of two planar pieces.

\begin{figure}[t]
  \centering
  \begin{minipage}[b]{0.30\linewidth}
    \centering
    \includegraphics[
      height=4.5cm
    ]{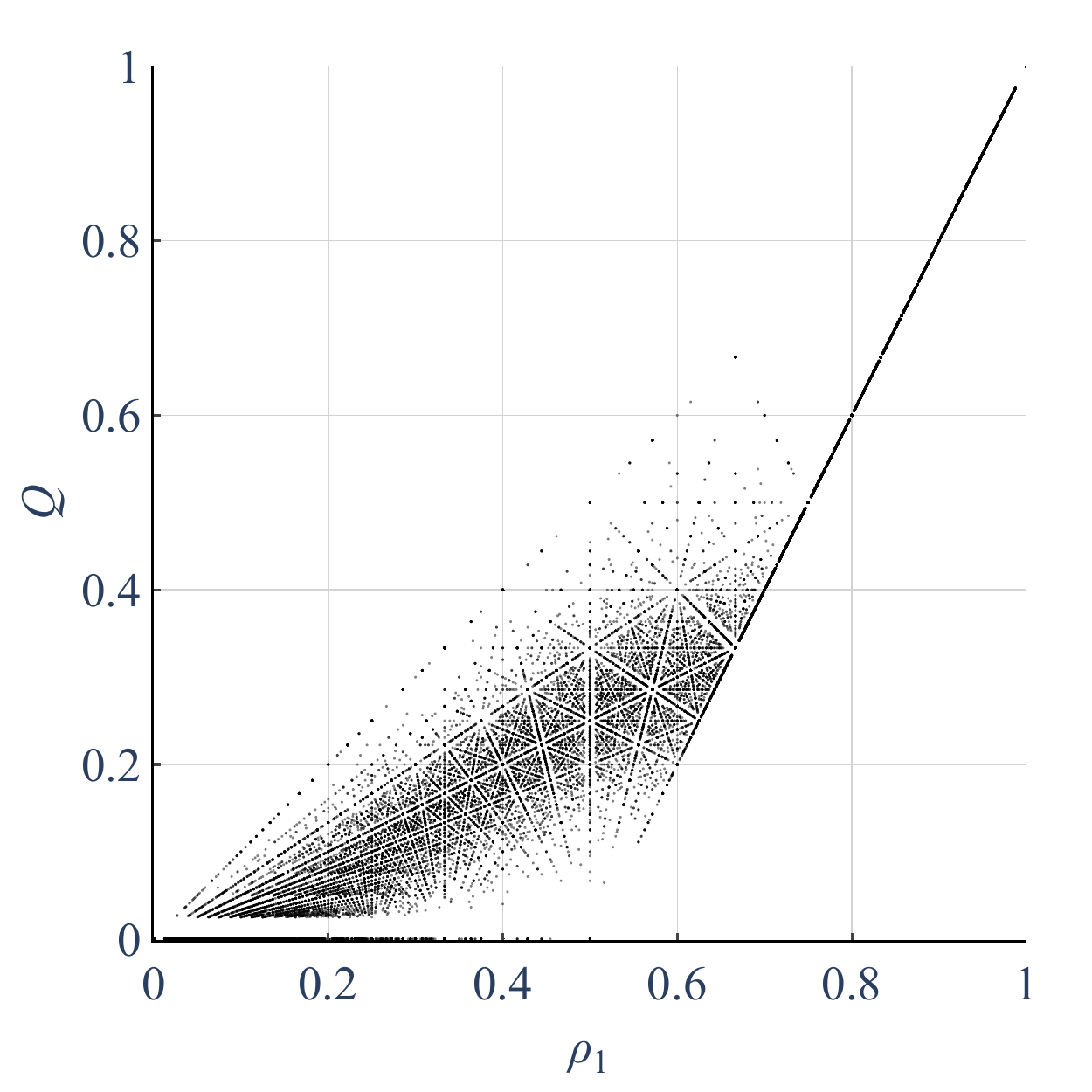}
    \subcaption{}
  \end{minipage}
  \hspace{-1.5em}
  \begin{minipage}[b]{0.34\linewidth}
    \centering
    \includegraphics[
      height=5.5cm,
      trim=1.5cm 0.8cm 1.2cm 0.4cm,
      clip
    ]{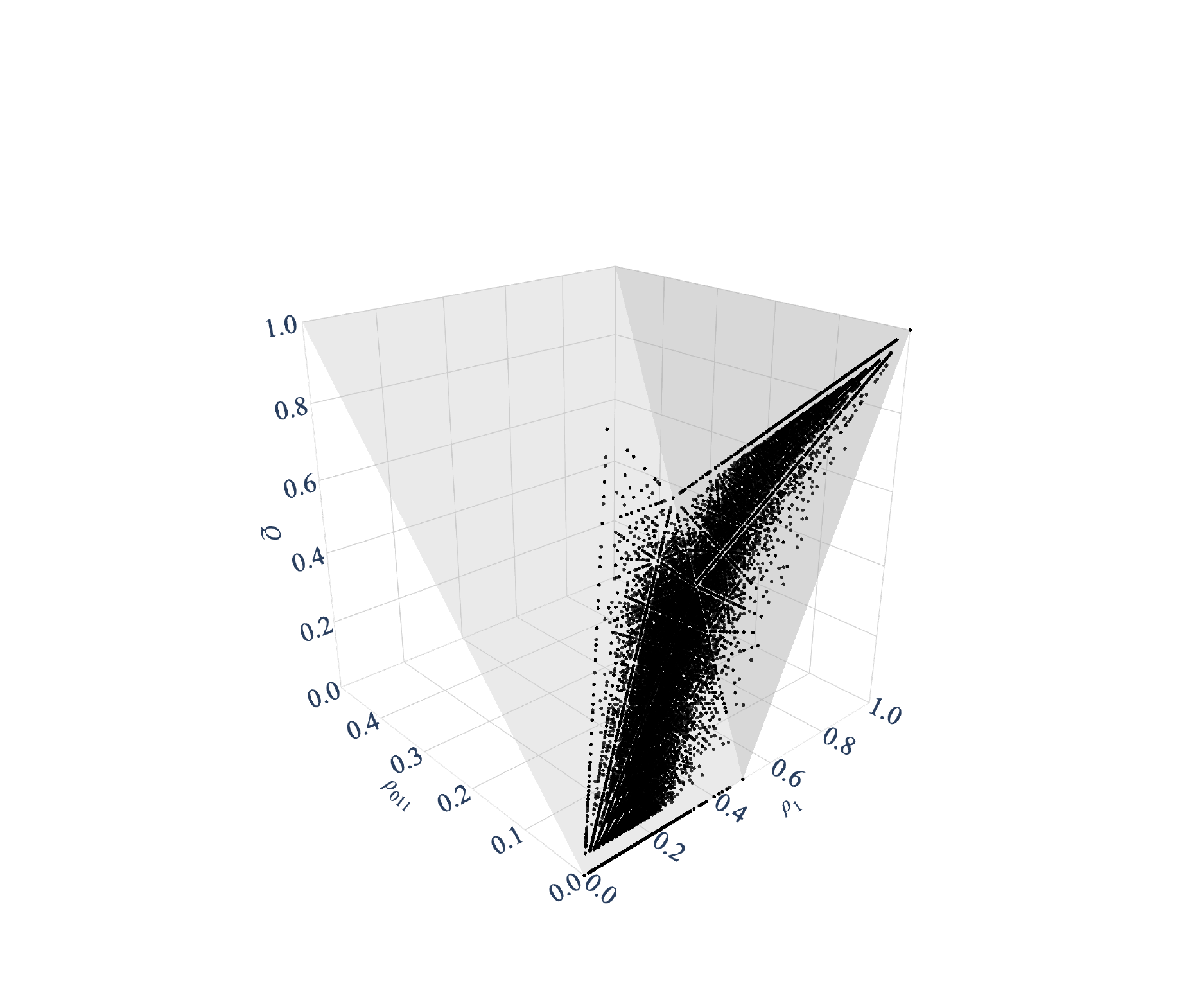}
    \subcaption{}
  \end{minipage}
  \hspace{-1.5em}
  \begin{minipage}[b]{0.34\linewidth}
    \centering
    \includegraphics[
      height=5.5cm,
      trim=1.5cm 0.8cm 1.2cm 0.8cm,
      clip
    ]{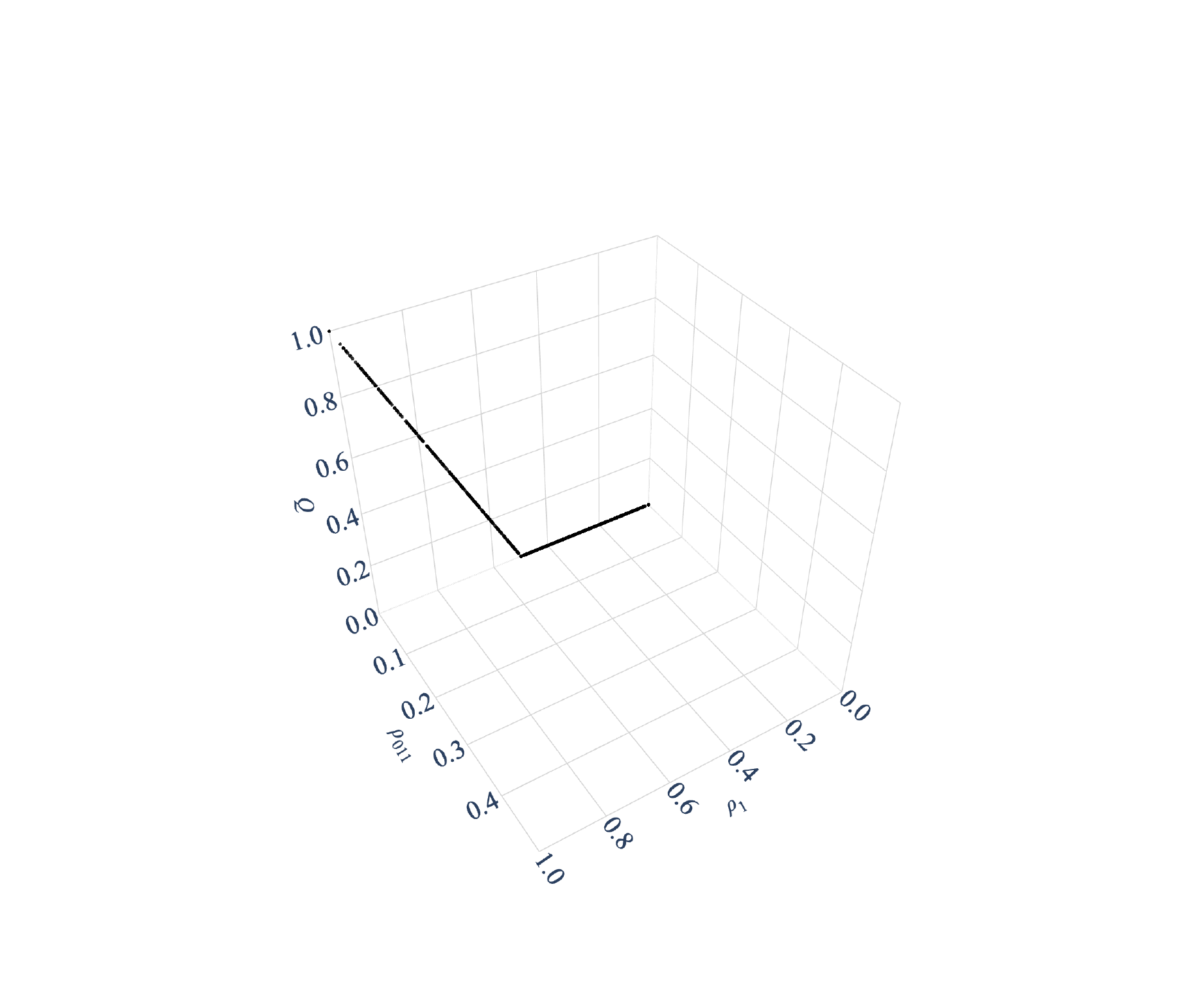}
    \subcaption{}
  \end{minipage}
  \caption{
  Fundamental diagrams of PCA5-17R.
  (a) Conventional two-dimensional fundamental diagram obtained by numerical simulations.
  (b) Three-dimensional fundamental diagram in the space $(\rho_1,\rho_{011},Q)$. The black points are numerical data, and the gray surface represents the theoretical formula $Q=\max(2\rho_1-1,2\rho_{011})$.
  (c) The same diagram viewed from an angle highlighting the two planar pieces $Q=2\rho_1-1$ and $Q=2\rho_{011}$.
  }
  \label{fig:pca5_17}
\end{figure}

Figure~\ref{fig:pca5_17} shows the fundamental diagrams of PCA5-17R. Figure~\ref{fig:pca5_17} (a) shows the conventional two-dimensional fundamental diagram in the plane $(\rho_1,Q)$. Figure~\ref{fig:pca5_17} (b) shows the three-dimensional fundamental diagram in the space $(\rho_1,\rho_{011},Q)$. Figure~\ref{fig:pca5_17} (c) shows the same diagram from a different viewpoint that highlights the two planar pieces corresponding to $Q=2\rho_1-1$ and $Q=2\rho_{011}$.

%%%%%%%%%%%%%%%%%%%%%%%%%%%%%%
\subsection{An example with an asymptotic second density}
%%%%%%%%%%%%%%%%%%%%%%%%%%%%%%
The result for PCA5-17R shows that a conserved second density can resolve the multivaluedness of the conventional density-flow relation. 
However, for some PCA5 rules, the additional density used for constructing a three-dimensional fundamental diagram is not conserved throughout the time evolution. 
Instead, numerical simulations indicate that it asymptotically approaches a constant value, and this asymptotic value can be used as the second coordinate of the diagram.

We present PCA5-28 as a representative example involving an asymptotic second density. The observed three-dimensional fundamental diagram of this rule consists of three planar pieces, in contrast to the two-plane structure of PCA5-17R. 
PCA5-28 is also written in the conservation form \eqref{eq:pca5_conservation_form}, where the flux function $q(a,b,c,d)$ is given in Table~\ref{tab:pca5_28_flux}.

\begin{table}[b]
\centering
\caption{Flux table of PCA5-28.}
\label{tab:pca5_28_flux}
\begin{tabular}{c|cccccccc}
$(a,b,c,d)$ & 1111 & 1110 & 1101 & 1100 & 1011 & 1010 & 1001 & 1000 \\ \hline
$q(a,b,c,d)$ & 0 & 1 & 1 & 1 & 0 & 1 & 0 & 1
\end{tabular}

\vspace{1mm}

\begin{tabular}{c|cccccccc}
$(a,b,c,d)$ & 0111 & 0110 & 0101 & 0100 & 0011 & 0010 & 0001 & 0000 \\ \hline
$q(a,b,c,d)$ & 0 & 0 & 1 & 1 & $-1$ & 0 & 0 & 0
\end{tabular}
\end{table}

\begin{figure}[t]
  \centering
  \begin{minipage}[b]{0.42\linewidth}
    \centering
    \includegraphics[
      height=6.5cm
    ]{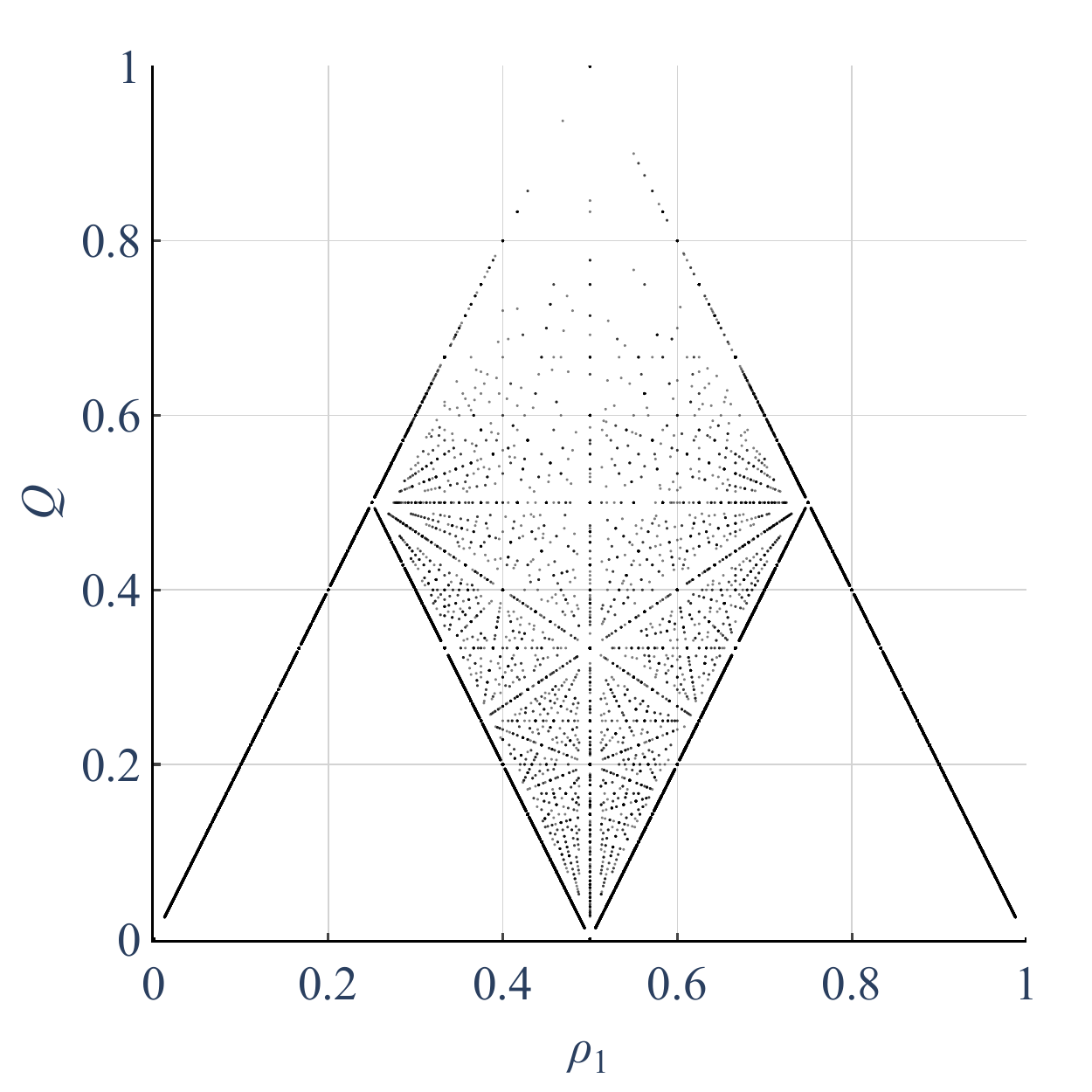}
    \subcaption{}
  \end{minipage}
  \hspace{1em}
  \begin{minipage}[b]{0.52\linewidth}
    \centering
    \includegraphics[
      height=7.5cm,
      trim=1.5cm 0.5cm 1.2cm 0.5cm,
      clip
    ]{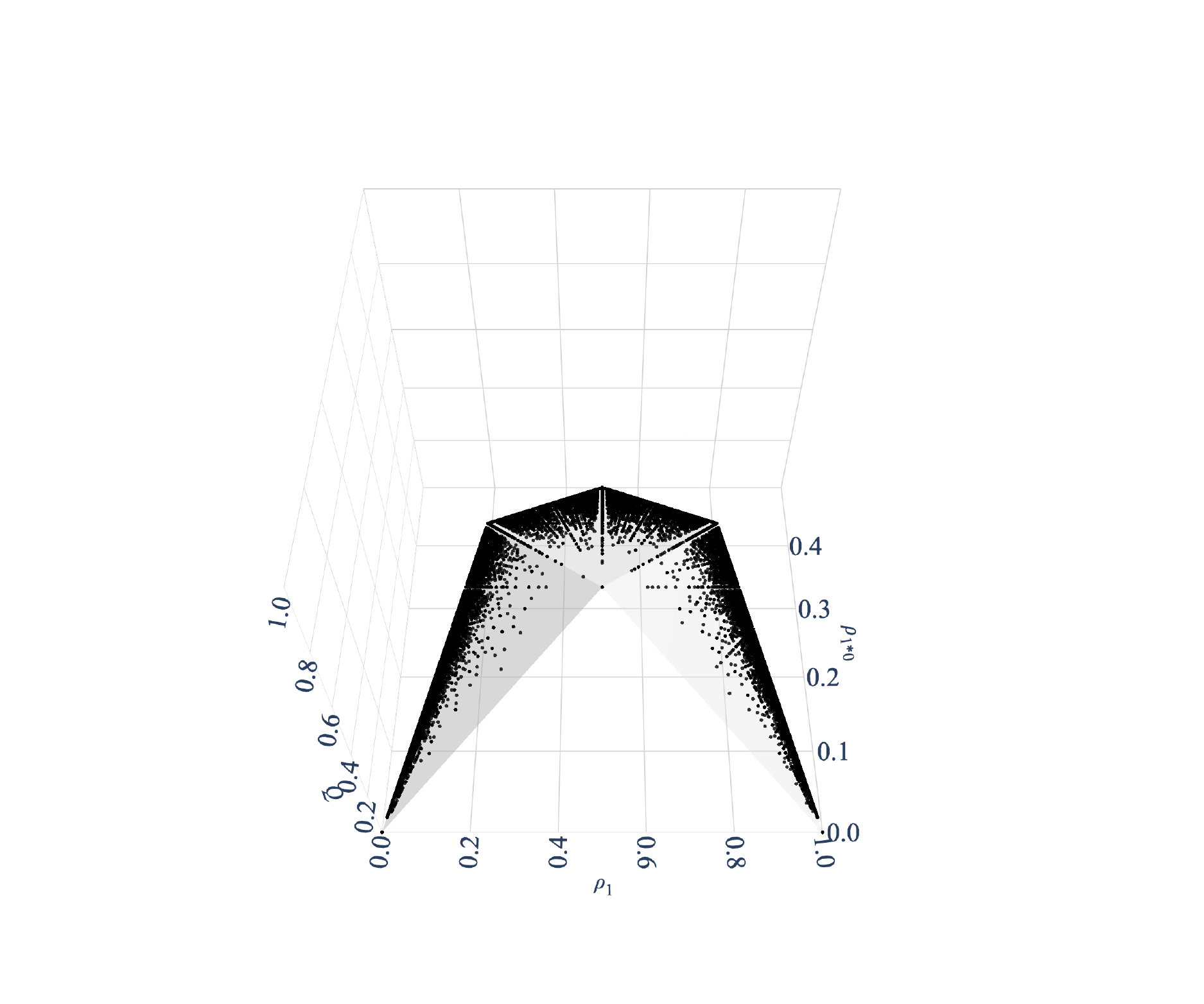}
    \subcaption{}
  \end{minipage}
  \caption{
  Fundamental diagrams of PCA5-28.
  (a) Conventional two-dimensional fundamental diagram obtained by numerical simulations.
  (b) Three-dimensional fundamental diagram in the space $(\rho_1,\rho_{1*0},Q)$. The black points are numerical data, and the gray surface represents the numerically observed relation
  $Q=\min(2\rho_1,-2\rho_{1*0}+1,-2\rho_1+2)$.
  }
  \label{fig:pca5_28}
\end{figure}

For PCA5-28, the conventional two-dimensional fundamental diagram in the plane $(\rho_1,Q)$ is multivalued. 
We use the asymptotic density $\rho_{1*0}$ of the local patterns $1*0$.
Here, the symbol $*$ means either 0 or 1, and we define
\begin{equation}
  \rho_{1*0}
  =
  \frac{\#100+\#110}{K}.
  \label{eq:rho_1star0}
\end{equation}
Numerical simulations indicate that the data in the space $(\rho_1,\rho_{1*0},Q)$ lie on a surface consisting of three planar pieces.
The numerically observed relation is
\begin{equation}
  Q
  =
  \min(2\rho_1,\,-2\rho_{1*0}+1,\,-2\rho_1+2).
  \label{eq:pca5_28_fd}
\end{equation}
Thus, although $\rho_1$ alone does not determine $Q$ uniquely, the pair $(\rho_1,\rho_{1*0})$ uniquely determines the observed mean flow. 
Since \eqref{eq:pca5_28_fd} is the minimum of three linear functions, the corresponding surface consists of three planar pieces. Figure~\ref{fig:pca5_28} shows the fundamental diagrams of PCA5-28. 
Figure~\ref{fig:pca5_28} (a) shows the conventional two-dimensional fundamental diagram in the plane $(\rho_1,Q)$. Figure~\ref{fig:pca5_28} (b) shows the three-dimensional fundamental diagram in the space $(\rho_1,\rho_{1*0},Q)$. 

The example of PCA5-28 suggests that the multivaluedness of the conventional two-dimensional fundamental diagram can be resolved by using a suitable second density evaluated in the asymptotic regime. Numerical simulations indicate that similar multi-planar three-dimensional fundamental diagrams are also observed for several other PCA5 rules.

The numerically observed examples are summarized in Table~\ref{tab:binary_pca5_summary}. Except for PCA5-17R, the listed equations of diagrams are derived by numerical simulations in the asymptotic regime, and have not yet been derived analytically. These numerically observed binary fundamental diagrams provide reference examples for the max-plus algebraic extensions studied in the following section.
\begin{table}[b]
\centering
\caption{Three-dimensional fundamental diagrams for binary PCA5 rules.}
\label{tab:binary_pca5_summary}
\begin{tabular}{l|l|l}
\hline
Rule & Second density & Three-dimensional fundamental diagram \\ \hline

PCA5-11R
& \qquad$\rho_{1*0}$
& $Q=\min(2\rho_1,\,-2\rho_{1*0}+1)$ \\[1mm]

PCA5-19R
& \qquad$\rho_{01*1}$
& $Q=\max(\min(2\rho_1,\,-2\rho_1+1), \, \rho_1-\rho_{01*1})$ \\[1mm]

PCA5-24R
& \qquad$\rho_{0*01}$
& $Q=\max(\min(2\rho_1,\,-\rho_1-\rho_{0*01}+1), \, 2\rho_1-1)$ \\[1mm]

PCA5-17R
& \qquad$\rho_{011}$
& $Q=\max(2\rho_{011},\,2\rho_1-1)$ \\[1mm]

PCA5-20
& \qquad$\rho_{1*10}$
& $Q=\min(2\rho_1,\,-\rho_1+\rho_{1*10}+1)$ \\[1mm]

PCA5-28
& \qquad$\rho_{1*0}$
& $Q=\min(2\rho_1,\,-2\rho_{1*0}+1,\,-2\rho_1+2)$ \\[1mm]

PCA5-42
& \qquad$\rho_{0*01}$
& $Q=\min\!\left(\max\!\left(\min(2\rho_1,\,-\rho_1-\rho_{0*01}+1),\,2\rho_1-1\right),\,-2\rho_1+2\right)$ \\[1mm]

PCA5-56
& \qquad$\rho_{01*1}$
& $Q=\min(\rho_1-\rho_{01*1},\,-2\rho_1+2)$ \\[1mm]

PCA5-57
& \qquad$\rho_{\mathrm{odd}}$
& $Q=\min(2(1-\rho_1),\,\rho_1-\rho_{\mathrm{odd}})$ \\[1mm]
PCA5-61
& \qquad$\rho_{01*0}$
& $Q=\min(-2\rho_1-\rho_{01*0}+2,\,\rho_1-\rho_{01*0})$ \\[1mm]

PCA5-93
& \qquad$\rho_{01*0}$
& $Q=\min(\rho_1-\rho_{01*0},\,1-\rho_1)$ \\ \hline

\end{tabular}
\end{table}
%%%%%%%%%%%%%%%%%%%%%%%%%%%%%%
\section{Real-valued max-plus extensions of PCA5}
\label{sec:maxplus_extensions}
%%%%%%%%%%%%%%%%%%%%%%%%%%%%%%
In the previous section, we considered three-dimensional fundamental diagrams for binary PCA5 rules. 
In this section, we discuss real-valued extensions of PCA5-17R and PCA5-93 using max-plus algebra. 
Our aim is to construct extensions that reproduce the original binary dynamics when the state variable is restricted to $\{0,1\}$, preserving the single-valued structure of the three-dimensional fundamental diagram in the real-valued setting.

Initial configurations used here can take real values, but we used random initial configurations in $[0,1]^K$ with rational components instead of arbitrary real values to avoid the numerical error. For each sample, the system size $K$ was selected from $5,\ldots,80$, and a denominator $m$ of rational value was selected from $1,\ldots,11$, both uniformly at random. 
Each component of the initial configuration was then chosen independently from $\left\{0,1/m,2/m,\ldots,m/m\right\}$.
Each sample was evolved for $1000$ transient steps. 
The mean flow $Q$ and the second density $\rho_2$ were then evaluated in the asymptotic regime.

As in the binary case, the real-valued extensions can be written in the conservation form \eqref{eq:pca5_conservation_form}. 
The difference is that the state variable is allowed to take real values and the flux function $q$ is given by a max-plus expression. 
In this section, we mainly consider initial data satisfying $0\leq u_j^0\leq 1$ for all $j$.

A desirable extension should satisfy two requirements. 
First, when the state variable is restricted to $\{0,1\}$, the extended equation should reduce to the original binary PCA. 
Second, the additional density used for the three-dimensional fundamental diagram should also reduce to the corresponding binary pattern density in the binary case.

It should be emphasized that such an extension is not determined uniquely from the binary rule. 
Even if the extended flux and the extended second density agree with the binary flux and the binary pattern density on $\{0,1\}$, the resulting real-valued system need not have a single-valued three-dimensional fundamental diagram. 
Thus, an appropriate combination of the flux function and the second density is essential.
%%%%%%%%%%%%%%%%%%%%%%%%%%%%%%
\subsection{Max-plus extensions of PCA5-17R}
\label{subsec:maxplus_pca5_17}
%%%%%%%%%%%%%%%%%%%%%%%%%%%%%%
We introduce two max-plus extensions of PCA5-17R, denoted by mPCA5-17R-1 and mPCA5-17R-2. 
The labels ``-1'' and ``-2'' do not mean different binary PCA rules; rather, they distinguish two different real-valued extensions of the same binary rule. 
Both extensions reduce to PCA5-17R when the state variable is restricted to $\{0,1\}$, but they use different max-plus expressions for the flux $q$ and for the second density. 
Both systems are written in the conservation form \eqref{eq:pca5_conservation_form}.

For both systems, the particle density
\begin{equation}
  \rho_1
  =
  \frac{1}{K}\sum_{j=1}^{K}u_j^n
  \label{eq:maxplus_rho1}
\end{equation}
is conserved because the time evolution is in the conservation form.

For PCA5-17R, the second coordinate of the three-dimensional fundamental diagram in the binary case is the density $\rho_{011}$. 
In the max-plus extensions considered here, this pattern density is replaced by the spatial average of a real-valued local function. 
More precisely, for each extension, we define
\begin{equation}
  \rho_2
  =
  \frac{1}{K}\sum_{j=1}^{K}
  r(u_{j-1}^n,u_j^n,u_{j+1}^n),
  \label{eq:maxplus_rho2_general}
\end{equation}
where the local function $r$ is chosen separately for mPCA5-17R-1 and mPCA5-17R-2.

When the state variable is restricted to $\{0,1\}$, the function $r$ reduces to the indicator counting the local pattern $011$. 
Thus, $\rho_2$ coincides with the second density used for binary PCA5-17R. 
In the real-valued case, however, $\rho_2$ is not necessarily conserved during the time evolution. 
In the numerical computations below, we use the asymptotic value of $\rho_2$ as the second coordinate of the three-dimensional fundamental diagram.
%%%%%%%%%%%%%%%%%%%%%%%%%%%%%%
\subsubsection{mPCA5-17R-1}
%%%%%%%%%%%%%%%%%%%%%%%%%%%%%%
The first extension of PCA5-17R is defined by the flux
\begin{equation}
  q(a,b,c,d)
  =
  \max(0,\,a+b-1,\,b+c-a-d-1).
  \label{eq:pca5_17_1_flux}
\end{equation}
The second density is defined by
\begin{equation}
  \rho_2
  =
  \frac{1}{K}\sum_{j=1}^{K}
  r(u_{j-1}^n,u_j^n,u_{j+1}^n),
\qquad
  r(a,b,c)
  =
  \max(0,\,a+b-c-1).
  \label{eq:pca5_17_1_rho2}
\end{equation}
If $a,b,c\in\{0,1\}$, then $r(a,b,c)=1$ holds only for $(a,b,c)=(1,1,0)$, and $r(a,b,c)=0$ otherwise. Thus, in the binary case, $\rho_2$ coincides with the density of the local pattern $110$, that is, $\rho_{110}$.
For this system, the interval $[0,1]$ is invariant under the time evolution. 
That is, if $0\leq u_j^n\leq 1$ for all $j$, then the same inequality holds for $u_j^{n+1}$ for all $j$. 
Therefore, the system gives a real-valued extension of the binary PCA5-17R in the state space $[0,1]^K$.

\begin{figure}[t]
  \centering
  \begin{minipage}[b]{0.42\linewidth}
    \centering
    \includegraphics[
      height=7.5cm,
      trim=1.5cm 0.5cm 1.2cm 0.5cm,
      clip
    ]{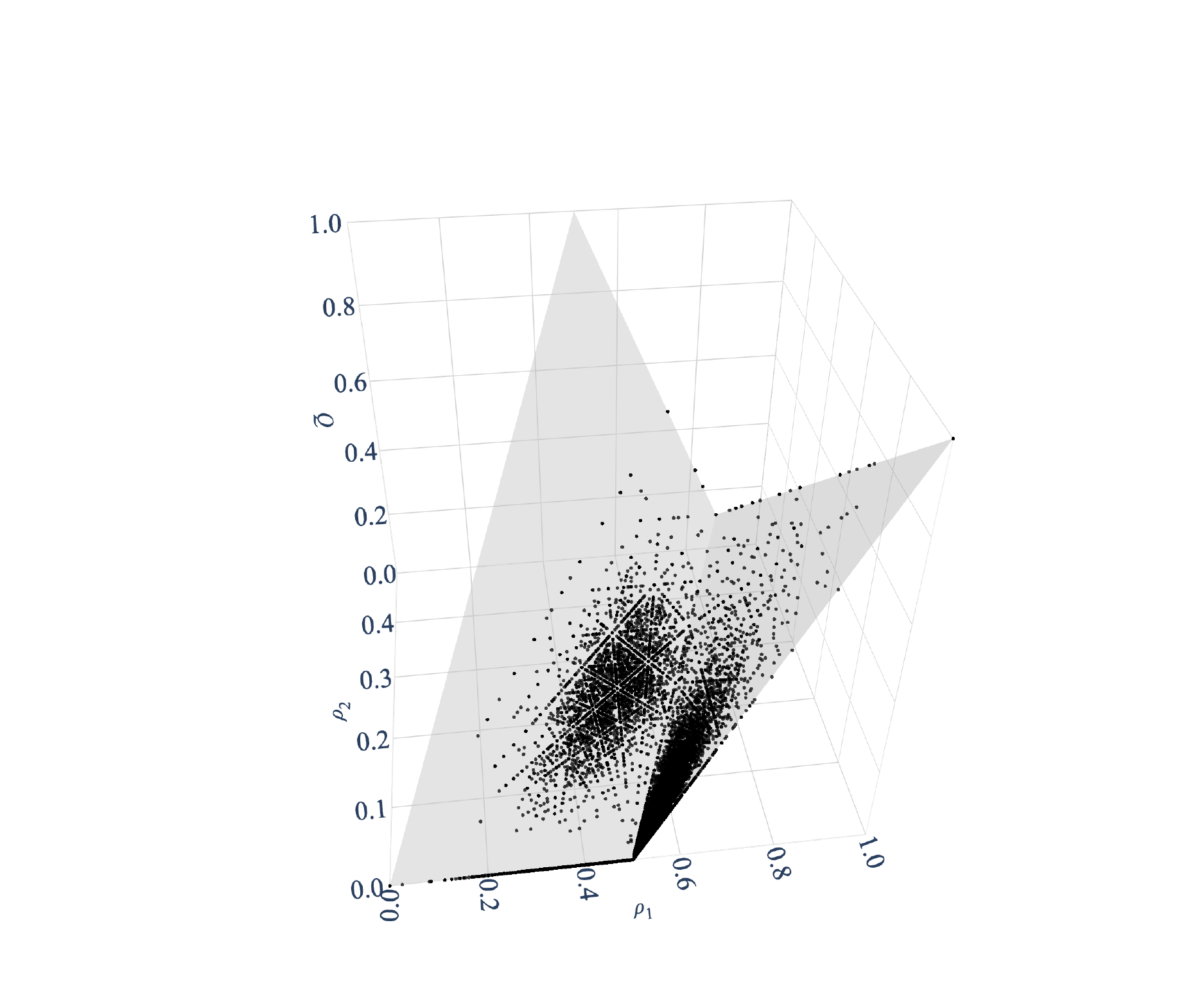}
    \subcaption{}
  \end{minipage}
  \hspace{1em}
  \begin{minipage}[b]{0.52\linewidth}
    \centering
    \includegraphics[
      height=7.5cm,
      trim=1.5cm 1.5cm 1.2cm 1.5cm,
      clip
    ]{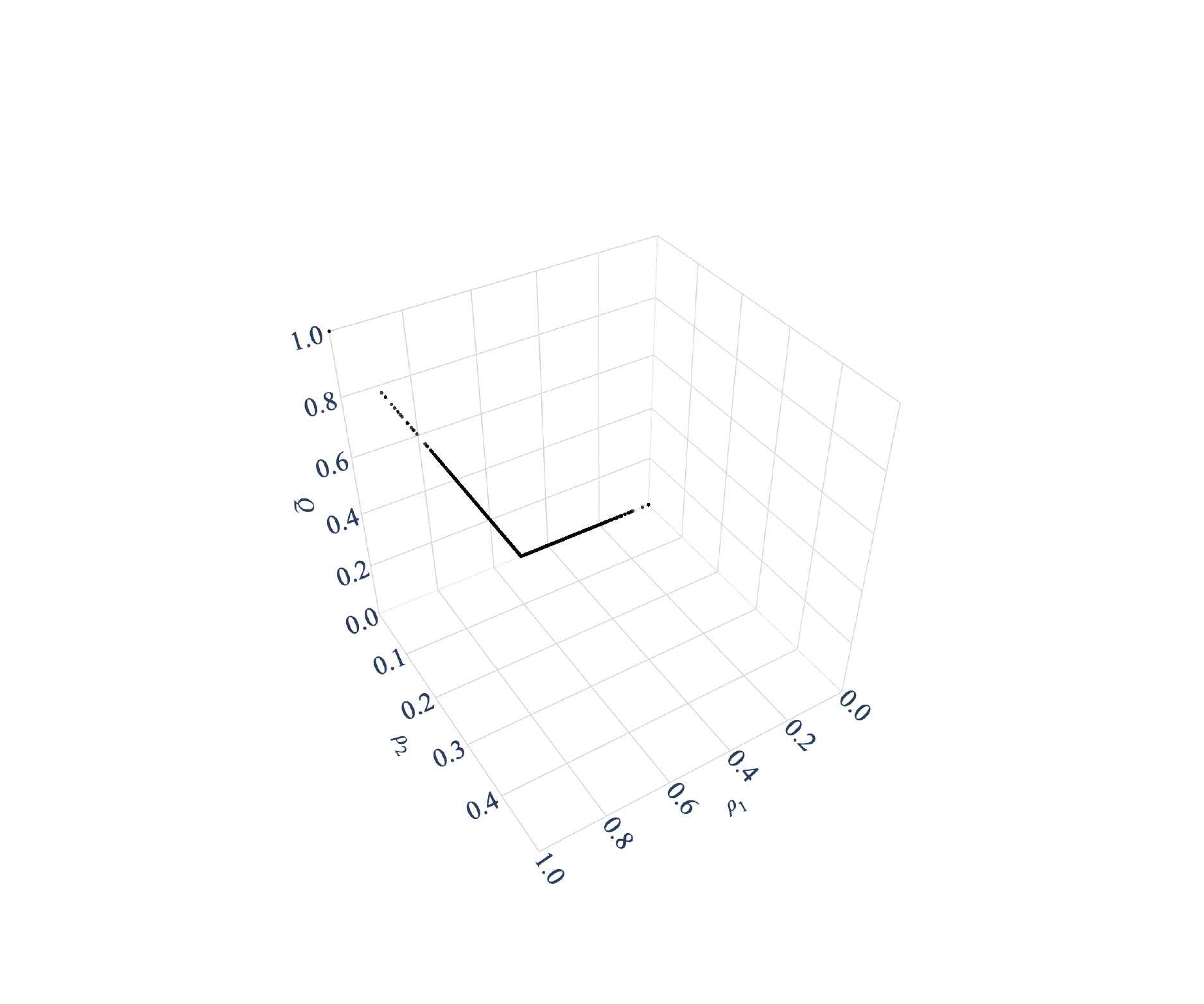}
    \subcaption{}
  \end{minipage}
  \caption{
  Three-dimensional fundamental diagram of the max-plus extension mPCA5-17R-1.
  (a) The diagram in the space $(\rho_1,\rho_2,Q)$.
  (b) The same diagram viewed from an angle highlighting the two planar pieces.
  The black points are numerical data, and the gray surface represents
  $Q=\max(2\rho_2,2\rho_1-1)$.
  }
  \label{fig:pca5_17_1_maxplus_fd}
\end{figure}

Numerical simulations show that $\rho_2$ asymptotically converges to a constant value.
Using this asymptotic value of $\rho_2$, the numerical data lie on a single-valued surface in the space $(\rho_1,\rho_2,Q)$. 
The observed three-dimensional fundamental diagram is
\begin{equation}
  Q
  =
  \max(2\rho_2,\,2\rho_1-1).
  \label{eq:pca5_17_1_3dfd}
\end{equation}
Figure~\ref{fig:pca5_17_1_maxplus_fd} (a) shows the three-dimensional fundamental diagram of mPCA5-17R-1 in the space $(\rho_1,\rho_2,Q)$. 
Figure~\ref{fig:pca5_17_1_maxplus_fd} (b) shows the same diagram from another angle, which highlights the two planar pieces corresponding to $Q=2\rho_2$ and $Q=2\rho_1-1$.

%%%%%%%%%%%%%%%%%%%%%%%%%%%%%%
\subsubsection{mPCA5-17R-2}
%%%%%%%%%%%%%%%%%%%%%%%%%%%%%%
The second extension of PCA5-17R is defined by another choice of the flux and the second density. 
The flux is given by
\begin{equation}
  q(a,b,c,d)
  =
  \max\big(
    a+b-1,\,
    \min(\max(0,a+b-1),1-c)
    +
    \min(\max(0,b+c-1),1-d)
  \big).
  \label{eq:pca5_17_2_flux}
\end{equation}
The corresponding local function for the second density is
\begin{equation}
  \rho_2
  =
  \frac{1}{K}\sum_{j=1}^{K}
  r(u_{j-1}^n,u_j^n,u_{j+1}^n),
\qquad
  r(a,b,c)
  =
  \min(\max(0,a+b-1),1-c).
  \label{eq:pca5_17_2_rho2}
\end{equation}
The form of mPCA5-17R-2 was suggested by J. Matsukidaira in private communication \cite{matsukidaira2025}.
As in the case of mPCA5-17R-1, the function $r(a,b,c)$ reduces to the indicator counting the local pattern $110$ for $a$, $b$, $c\in\{0,1\}$. 
Hence, the second density agrees with $\rho_{110}$ in the binary restriction.

The system also preserves the interval $[0,1]$ under time evolution; that is, if $0\leq u_j^n\leq 1$ for all $j$, then $0\leq u_j^{n+1}\leq 1$ for all $j$. 
Thus, mPCA5-17R-2 is another real-valued max-plus extension of the same binary PCA5-17R rule. 
Although the expressions \eqref{eq:pca5_17_1_flux}, \eqref{eq:pca5_17_1_rho2} and \eqref{eq:pca5_17_2_flux}, \eqref{eq:pca5_17_2_rho2} are different, the numerical results show that the two systems give rise to the same three-dimensional fundamental diagram. For mPCA5-17R-2, numerical simulations indicate that the asymptotic value of $\rho_2$ together with $\rho_1$ determines the mean flow $Q$. 
The observed three-dimensional fundamental diagram is again
\begin{equation}
  Q
  =
  \max(2\rho_2,\,2\rho_1-1).
  \label{eq:pca5_17_2_3dfd}
\end{equation}

\begin{figure}[t]
  \centering
  \begin{minipage}[b]{0.42\linewidth}
    \centering
    \includegraphics[
      height=7.5cm,
      trim=1.5cm 0.5cm 1.2cm 0.5cm,
      clip
    ]{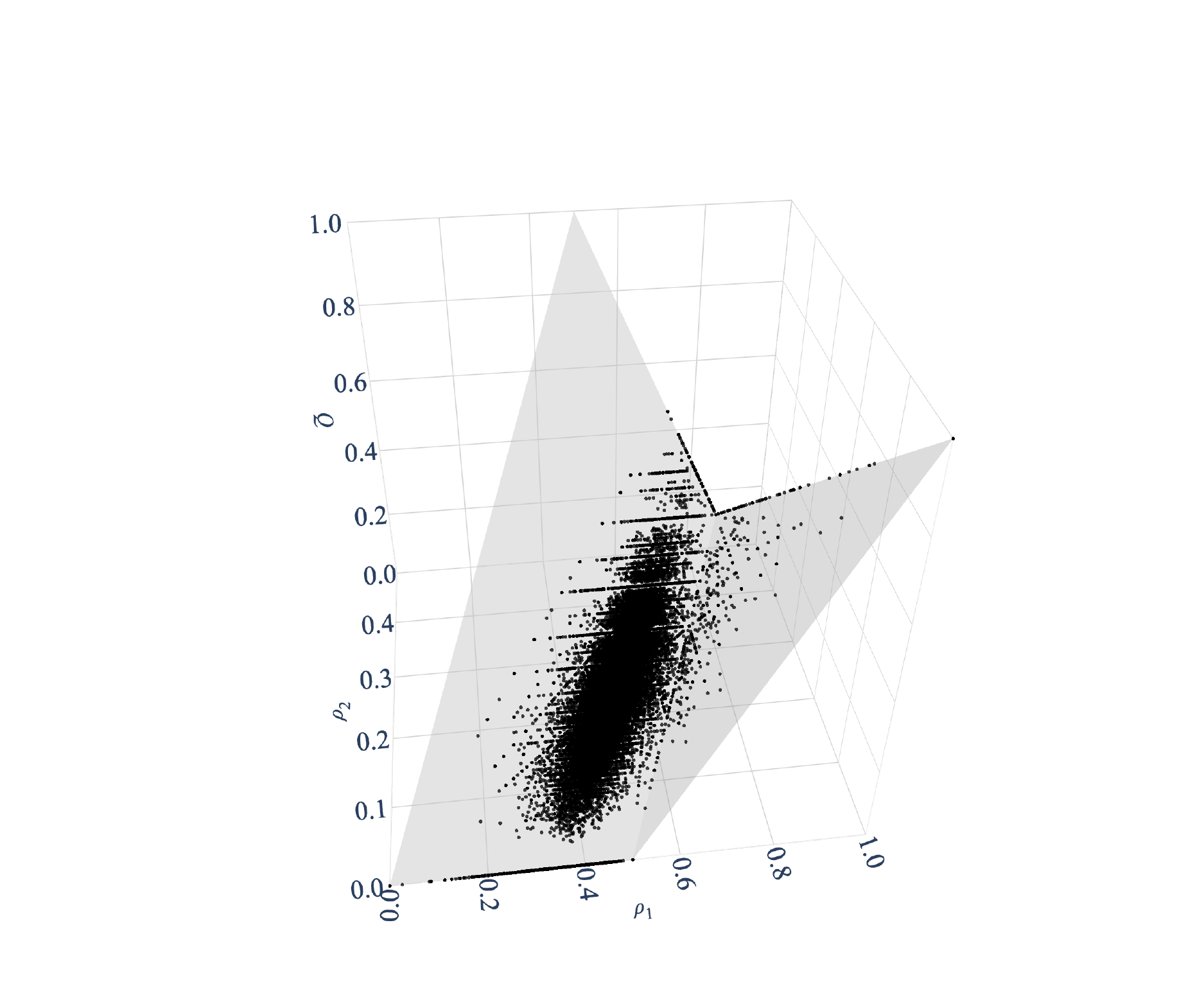}
    \subcaption{}
  \end{minipage}
  \hspace{1em}
  \begin{minipage}[b]{0.52\linewidth}
    \centering
    \includegraphics[
      height=7.5cm,
      trim=1.5cm 1.5cm 1.2cm 1.5cm,
      clip
    ]{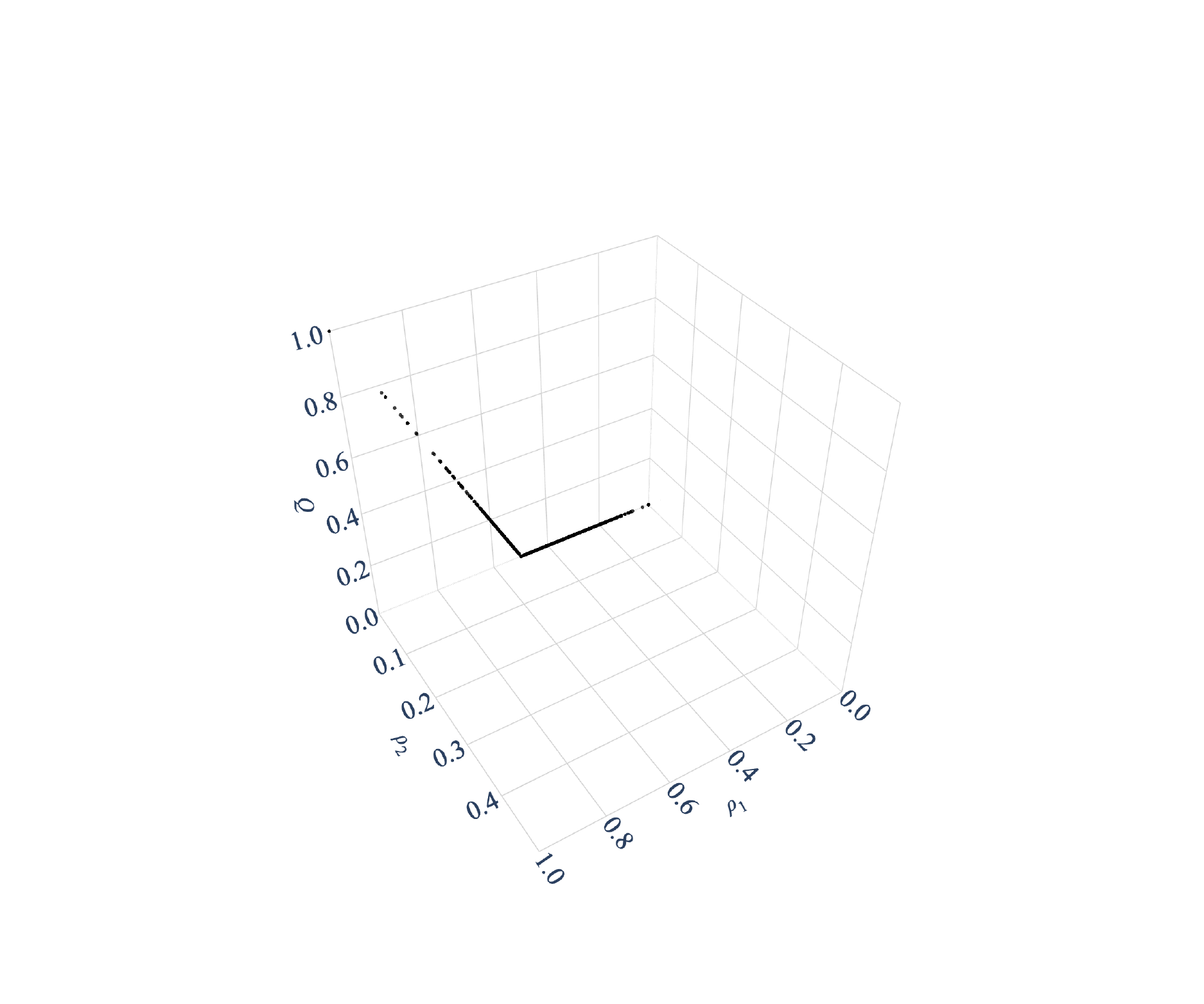}
    \subcaption{}
  \end{minipage}
  \caption{
  Three-dimensional fundamental diagram of the max-plus extension mPCA5-17R-2.
  (a) The diagram in the space $(\rho_1,\rho_2,Q)$.
  (b) The same diagram viewed from an angle highlighting the two planar pieces.
  The black points are numerical data, and the gray surface represents
  $Q=\max(2\rho_2,2\rho_1-1)$.
  }
  \label{fig:pca5_17_2_maxplus_fd}
\end{figure}

Figure~\ref{fig:pca5_17_2_maxplus_fd} (a) shows the three-dimensional fundamental diagram of mPCA5-17R-2 in the space $(\rho_1,\rho_2,Q)$. 
Figure~\ref{fig:pca5_17_2_maxplus_fd} (b) shows the same diagram from another angle, which highlights the two planar pieces corresponding to $Q=2\rho_2$ and $Q=2\rho_1-1$. 
As in mPCA5-17R-1, the numerical data lie on the surface given by \eqref{eq:pca5_17_2_3dfd}. 
Thus, although mPCA5-17R-1 and mPCA5-17R-2 have different forms, they lead to the same macroscopic relation. We have not succeeded in obtaining a complete classification of the asymptotic states of mPCA5-17R-1 and mPCA5-17R-2, though the numerical results in Figures~\ref{fig:pca5_17_1_maxplus_fd} and \ref{fig:pca5_17_2_maxplus_fd} support the macroscopic relations \eqref{eq:pca5_17_1_3dfd} and \eqref{eq:pca5_17_2_3dfd}.
%%%%%%%%%%%%%%%%%%%%%%%%%%%%%%
\subsection{Other max-plus extensions}
\label{subsec:other_maxplus_extensions}
%%%%%%%%%%%%%%%%%%%%%%%%%%%%%%
The above two systems show that different max-plus extensions of the same binary PCA5 rule can lead to the same three-dimensional fundamental diagram. 
We next summarize the corresponding results for PCA5-93, whose second density is not a conserved density in the binary setting. 
The relations are supported by numerical simulations in the asymptotic regime.

For PCA5-93, we considered two max-plus extensions, denoted by mPCA5-93-1 and mPCA5-93-2. 
In the binary case, the second density is the density of the local pattern $01*0$, where $*$ denotes either 0 or 1. 
This density is not conserved during the time evolution, and its asymptotic value is used as the second coordinate of the three-dimensional fundamental diagram.

Both systems are written in the conservation form \eqref{eq:pca5_conservation_form}. 
For mPCA5-93-1, the flux and the local function defining the second density are given by
\begin{equation}
  q(a,b,c,d)
  =
  \min(a+d,\,b,\,1-c),
  \qquad
  r(a,b,c,d)
  =
  \max(0,\,b-a-d).
  \label{eq:pca5_93_1_flux_r}
\end{equation}
For mPCA5-93-2, they are given by
\begin{equation}
  q(a,b,c,d)
  =
  \min(\max(a,d),\,b,\,1-c),
  \qquad
  r(a,b,c,d)
  =
  \max(0,\,b-\max(a,d)).
  \label{eq:pca5_93_2_flux_r}
\end{equation}
In both cases, the second density is defined by
\begin{equation}
  \rho_2
  =
  \frac{1}{K}\sum_{j=1}^{K}
  r(u_{j-1}^n,u_j^n,u_{j+1}^n,u_{j+2}^n).
  \label{eq:pca5_93_rho2}
\end{equation}
When the state variables are restricted to $\{0,1\}$, the local function $r$ counts the local pattern $01*0$, and hence $\rho_2$ coincides with the binary density $\rho_{01*0}$. 
In the real-valued case, $\rho_2$ is not necessarily conserved. 
In the numerical calculation, we use the asymptotic value of $\rho_2$ as the second coordinate of the three-dimensional fundamental diagram.

For both mPCA5-93-1 and mPCA5-93-2, numerical simulations indicate that the data lie on the surface
\begin{equation}
  Q
  =
  \min(\rho_1-\rho_2,\,1-\rho_1).
  \label{eq:pca5_93_fd}
\end{equation}
This relation coincides with the binary three-dimensional fundamental diagram of PCA5-93. 
Thus, as in the case of PCA5-17R, two different max-plus extensions of PCA5-93 lead to the same macroscopic relation in the space $(\rho_1,\rho_2,Q)$.

These examples imply that the single-valued structure of a binary three-dimensional fundamental diagram can be preserved under suitable max-plus extensions. 
At the same time, the construction is not merely a formal extension of the binary flux; the flux and the second density must be chosen consistently.

\begin{table}[t]
\centering
\caption{Max-plus extensions of selected PCA5 rules.}
\label{tab:maxplus_extensions_summary}
\begin{tabular}{l|l|l}
\hline
Base rule & Extension & Observed relation \\ \hline
PCA5-17R
& mPCA5-17R-1, mPCA5-17R-2
& $Q=\max(2\rho_2,\,2\rho_1-1)$ \\[1mm]

PCA5-93
& mPCA5-93-1, mPCA5-93-2
& $Q=\min(\rho_1-\rho_2,\,1-\rho_1)$ \\
\hline
\end{tabular}
\end{table}
%%%%%%%%%%%%%%%%%%%%%%%%%%%%%%
\subsection{A remark on PCA5-29}
\label{subsec:pca5_29_remark}
%%%%%%%%%%%%%%%%%%%%%%%%%%%%%%
Finally, we briefly mention PCA5-29 as another example of a real-valued max-plus extension. 
This rule is not analyzed in detail in this paper, but it is noteworthy because its numerically observed three-dimensional fundamental diagram has a different geometric structure from those of PCA5-17R and PCA5-93. A max-plus extension of PCA5-29 is written in the conservation form \eqref{eq:pca5_conservation_form}, where the flux function is given by
\begin{equation}
  q(a,b,c,d)
  =
  \min(b,\,1-c)
  +
  \max\left(0,\,\min(a,\,1-b-\min(c,d))\right).
  \label{eq:pca5_29_flux}
\end{equation}
The second density is defined by
\begin{equation}
  \rho_2
  =
  \frac{1}{K}\sum_{j=1}^{K}
  \min(u_{j-1}^n,\,1-u_j^n).
  \label{eq:pca5_29_rho2}
\end{equation}

Numerical simulations suggest that the corresponding three-dimensional fundamental diagram consists of two planar pieces. 
One component lies on the plane $\rho_1=\rho_2$, whereas the other lies on the plane $Q=1-\rho_1$.
This example is different from the preceding cases since the former planar component is vertical with respect to the plane $(\rho_1,\rho_2)$. Therefore, the usual treatment to represent the mean flow $Q$ may need to be reconsidered for such systems.

In this paper, PCA5-29 is left as a subject for future work. 
A detailed analysis of its asymptotic behavior, the precise role of the second density, and a theoretical derivation of the three-dimensional fundamental diagram remain open.
%%%%%%%%%%%%%%%%%%%%%%%%%%%%%%
\section{Conclusion}
\label{sec:conclusion}
%%%%%%%%%%%%%%%%%%%%%%%%%%%%%%
In this paper, we studied three-dimensional fundamental diagrams for binary PCA5 rules and their real-valued max-plus extensions. 
The main purpose was to examine how the multivaluedness of the conventional density-flow relation can be resolved by introducing a second density, and whether the resulting single-valued structure can be preserved in real-valued max-plus extensions.

For binary PCA5 rules, we first reviewed PCA5-17R as a case in which the second density is a conserved density. 
In this rule, the mean flow is uniquely determined by the particle density $\rho_1$ and the additional conserved density $\rho_{011}$. 
We then considered PCA5-28 as a representative example involving an asymptotic second density. 
Numerical simulations indicate that the pair of $\rho_1$ and the asymptotic value $\rho_{1*0}$ determines a single-valued three-dimensional fundamental diagram. 
The additional examples summarized in Table~\ref{tab:binary_pca5_summary} show that similar multi-planar three-dimensional fundamental diagrams are observed for several other PCA5 rules.

We next introduced real-valued max-plus extensions of PCA5-17R. 
Two different extensions, mPCA5-17R-1 and mPCA5-17R-2, were considered. 
Although their fluxes and local functions defining the second density are different, both systems reduce to the same binary PCA5-17R rule when the state variable is restricted to $\{0,1\}$. 
Numerical simulations indicate that, in both extensions, the asymptotic value of the extended second density $\rho_2$ leads to the same three-dimensional fundamental diagram. 
This shows that different real-valued extensions of the same binary rule can preserve the same macroscopic relation.

We also summarized max-plus extensions of PCA5-93. 
As in the case of PCA5-17R, two different extensions, mPCA5-93-1 and mPCA5-93-2, were considered. 
Although the two extensions use different fluxes and different local functions defining the second density, numerical simulations indicate that both systems lead to the same three-dimensional relation as the corresponding binary PCA5-93.

The above results show that the single-valued structure of a binary three-dimensional fundamental diagram can be preserved under suitable real-valued max-plus extensions. 
At the same time, the construction is not merely a formal extension of the binary flux function; the flux function and the second density must be chosen in a compatible way.

Finally, we mentioned PCA5-29 as a further example. 
Numerical simulations suggest that its three-dimensional fundamental diagram has a different geometric structure, including a planar component that is vertical with respect to the plane $(\rho_1,\rho_2)$. 
This indicates that the usual description of mean flow $Q$ as a single-valued function of $(\rho_1,\rho_2)$ may need to be refined for some max-plus extended systems.

Several problems remain open. 
First, the relations among PCA5 rules giving the single-valued three-dimensional fundamental diagrams and their general structure remain to be clarified. Second, theoretical derivation of the three-dimensional fundamental diagrams for the max-plus extensions of PCA5-17R and PCA5-93 should be developed. 
Third, it is important to clarify systematic criteria for choosing the flux and the second density so that a real-valued max-plus extension preserves the single-valued structure of the binary three-dimensional fundamental diagram. 
These problems are expected to provide a deeper understanding of the relations among binary particle cellular automata, max-plus algebraic extensions, and fundamental diagrams.
%%%%%%%%%%%%%%%%%%%%%%%%%%%%%%
\appendix
\section{Rules of PCA5}
\label{app:pca5_rule_table}
%%%%%%%%%%%%%%%%%%%%%%%%%%%%%%
This appendix lists the rules of PCA5. The first column denotes the sequential group number of PCA5. The second column gives the canonical representative of each PCA5 group. The third, fourth, fifth columns denote Boolean conjugation, reflection, and conjugate reflection of the canonical rule, respectively.

\scriptsize
\setlength{\tabcolsep}{2pt}
\renewcommand{\arraystretch}{1.15}

\begin{longtable}{
  c
  >{\centering\arraybackslash}p{0.22\linewidth}
  >{\centering\arraybackslash}p{0.22\linewidth}
  >{\centering\arraybackslash}p{0.22\linewidth}
  >{\centering\arraybackslash}p{0.22\linewidth}
}
\caption{Rules of PCA5.}
\label{tab:pca5_rules}
\\
\toprule
group & canonical rule & conjugation & reflection & conjugate reflection \\
\midrule
\endfirsthead

\caption[]{Rules of PCA5, continued.}
\\
\toprule
group & canonical rule & conjugation & reflection & conjugate reflection \\
\midrule
\endhead

\midrule
\multicolumn{5}{r}{Continued on the next page}
\\
\endfoot

\bottomrule
\endlastfoot

5-1 &
\texttt{1010101010101011}\newline\texttt{1010101010101000} &
\texttt{1110101010101010}\newline\texttt{0010101010101010} &
\texttt{1111111111111110}\newline\texttt{0000000000000010} &
\texttt{1011111111111111}\newline\texttt{1000000000000000} \\

5-2 &
\texttt{1010101110111000}\newline\texttt{1010100010111000} &
\texttt{1110001011101010}\newline\texttt{1110001000101010} &
\texttt{1111111111111000}\newline\texttt{0000000000111000} &
\texttt{1110001111111111}\newline\texttt{1110000000000000} \\

5-3 &
\texttt{1010101110111100}\newline\texttt{1010100010001100} &
\texttt{1100111011101010}\newline\texttt{1100001000101010} &
\texttt{1111111111101000}\newline\texttt{0000001100101000} &
\texttt{1110101100111111}\newline\texttt{1110100000000000} \\

5-4 &
\texttt{1010101110111110}\newline\texttt{1010100010000010} &
\texttt{1011111011101010}\newline\texttt{1000001000101010} &
\texttt{1111111011101011}\newline\texttt{0000001000101000} &
\texttt{1110101110111111}\newline\texttt{0010100010000000} \\

5-5 &
\texttt{1010101110111111}\newline\texttt{1010100010000000} &
\texttt{1111111011101010}\newline\texttt{0000001000101010} &
\texttt{1111111011101010}\newline\texttt{0000001000101010} &
\texttt{1010101110111111}\newline\texttt{1010100010000000} \\

5-6 &
\texttt{1010111110001000}\newline\texttt{1010110010111000} &
\texttt{1110001011001010}\newline\texttt{1110111000001010} &
\texttt{1111111111011000}\newline\texttt{0000110000011000} &
\texttt{1110011111001111}\newline\texttt{1110010000000000} \\

5-7 &
\texttt{1010111110001100}\newline\texttt{1010110010001100} &
\texttt{1100111011001010}\newline\texttt{1100111000001010} &
\texttt{1111111111001000}\newline\texttt{0000111100001000} &
\texttt{1110111100001111}\newline\texttt{1110110000000000} \\
\end{longtable}

%
% Each of the commands below will create an unnumbered section with the appropriate heading.
% Remove any sections that are not relevant for your article.
% All sections except suppdata will be removed if the [anonymous] option is used.
% See iopjournal-guidelines.pdf for more information.
%

%\ack{Sample text inserted for demonstration.}

%\funding{Sample text inserted for demonstration.}
% This section is a list of funder names and grant numbers

%\roles{Sample text inserted for demonstration.}
% List author names and the contributions made to the article, using terms from the NISO Contributor Roles Taxonomy (CRediT) https://credit.niso.org

%\data{Sample text inserted for demonstration.}
% For more information on IOP Publishing's research data policy see: https://publishingsupport.iopscience.iop.org/questions/research-data/

%\suppdata{Sample text inserted for demonstration.}
%%%%%%%%%%%%%%%%%%%%%%%%%%%%%%
\normalsize
\section{Fluxes of PCA5}
\label{app:pca5_flux_table}
%%%%%%%%%%%%%%%%%%%%%%%%%%%%%%
This appendix lists the fluxes of PCA5. The first column denotes the group number of PCA5. The second a flux of a canonical rule of PCA5. The sequence of the numbers denote the values of $q(1,1,1,1)$, $q(1,1,1,0)$, $q(1,1,0,1)$, $\ldots$, $q(0,0,0,0)$. The overlined numbers $\overline{1}$ and $\overline{2}$ denote $-1$ and $-2$, respectively. The third, fourth, fifth columns denote fluxes of Boolean conjugation, reflection, and conjugate reflection of the canonical, respectively.

\scriptsize
\setlength{\tabcolsep}{2pt}
\renewcommand{\arraystretch}{1.15}

\begin{longtable}{
  c
  >{\centering\arraybackslash}p{0.22\linewidth}
  >{\centering\arraybackslash}p{0.22\linewidth}
  >{\centering\arraybackslash}p{0.22\linewidth}
  >{\centering\arraybackslash}p{0.22\linewidth}
}
\caption{Fluxes of PCA5.}
\label{tab:pca5_fluxes}
\\
\toprule
group & canonical rule & conjugation & reflection & conjugate reflection \\
\midrule
\endfirsthead

\caption[]{Fluxes of PCA5, continued.}
\\
\toprule
group & canonical rule & conjugation & reflection & conjugate reflection \\
\midrule
\endhead

\midrule
\multicolumn{5}{r}{Continued on the next page}
\\
\endfoot

\bottomrule
\endlastfoot

5-1 &
$\overline{1}$\,0\,0\,1\,$\overline{1}$\,0\,0\,1\,$\overline{1}$\,0\,0\,1\,$\overline{1}$\,0\,0\,0
& $\overline{1}$\,$\overline{1}$\,$\overline{1}$\,0\,$\overline{2}$\,$\overline{1}$\,$\overline{1}$\,0\,$\overline{2}$\,$\overline{1}$\,$\overline{1}$\,0\,$\overline{2}$\,$\overline{1}$\,$\overline{1}$\,0
& 1\,1\,1\,1\,0\,0\,0\,0\,0\,0\,0\,0\,$\overline{1}$\,$\overline{1}$\,$\overline{1}$\,0
& 1\,2\,2\,2\,1\,1\,1\,1\,1\,1\,1\,1\,0\,0\,0\,0 \\
5-2 &
$\overline{1}$\,0\,0\,1\,$\overline{1}$\,0\,0\,0\,$\overline{1}$\,0\,0\,0\,$\overline{1}$\,0\,0\,0
& $\overline{1}$\,$\overline{1}$\,$\overline{1}$\,0\,$\overline{1}$\,$\overline{1}$\,$\overline{1}$\,0\,$\overline{1}$\,$\overline{1}$\,$\overline{1}$\,0\,$\overline{2}$\,$\overline{1}$\,$\overline{1}$\,0
& 1\,1\,1\,1\,0\,0\,0\,0\,0\,0\,0\,0\,$\overline{1}$\,0\,0\,0
& 1\,1\,1\,2\,1\,1\,1\,1\,1\,1\,1\,1\,0\,0\,0\,0 \\
5-3 &
$\overline{1}$\,0\,0\,1\,$\overline{1}$\,0\,0\,0\,$\overline{1}$\,0\,0\,0\,$\overline{1}$\,$\overline{1}$\,0\,0
& $\overline{1}$\,$\overline{1}$\,0\,0\,$\overline{1}$\,$\overline{1}$\,$\overline{1}$\,0\,$\overline{1}$\,$\overline{1}$\,$\overline{1}$\,0\,$\overline{2}$\,$\overline{1}$\,$\overline{1}$\,0
& 1\,1\,1\,1\,0\,0\,0\,0\,0\,0\,0\,1\,$\overline{1}$\,0\,0\,0
& 1\,1\,1\,2\,0\,1\,1\,1\,1\,1\,1\,1\,0\,0\,0\,0 \\
5-4 &
$\overline{1}$\,0\,0\,1\,$\overline{1}$\,0\,0\,0\,$\overline{1}$\,0\,0\,0\,$\overline{1}$\,$\overline{1}$\,$\overline{1}$\,0
& $\overline{1}$\,0\,0\,0\,$\overline{1}$\,$\overline{1}$\,$\overline{1}$\,0\,$\overline{1}$\,$\overline{1}$\,$\overline{1}$\,0\,$\overline{2}$\,$\overline{1}$\,$\overline{1}$\,0
& 1\,1\,1\,1\,0\,0\,0\,1\,0\,0\,0\,1\,$\overline{1}$\,0\,0\,0
& 1\,1\,1\,2\,0\,1\,1\,1\,0\,1\,1\,1\,0\,0\,0\,0 \\
\end{longtable}

\end{document}